\documentclass[12pt,preprint]{aastex}
\usepackage{natbib}
\def\lsim{\lower.5ex\hbox{$\; \buildrel < \over \sim \;$}}
\def\gsim{\lower.5ex\hbox{$\; \buildrel > \over \sim \;$}}

\begin{document}
\bibliographystyle{plainnat}

\title{Effects of Compton Cooling on Outflow in a Two Component Accretion Flow around 
a Black Hole: Results of a Coupled Monte Carlo-TVD Simulation}

\author{Sudip K. Garain\altaffilmark{1}, Himadri Ghosh\altaffilmark{1}, Sandip K. Chakrabarti\altaffilmark{2,1}}

\altaffiltext{1}{ S.N. Bose National Centre for Basic Sciences,
JD Block, Salt Lake, Sector III, Kolkata, 700098, \\ e-mail: sudip@bose.res.in}
\altaffiltext{1}{ S.N. Bose National Centre for Basic Sciences,
JD Block, Salt Lake, Sector III, Kolkata, 700098,\\ e-mail: himadri@bose.res.in}
\altaffiltext{2}{ Indian Centre for Space Physics, Chalantika 43, Garia Station Rd.,
Garia, Kolkata, 700084,\\ e-mail: chakraba@bose.res.in}

\begin{abstract}
We wish to investigate the effects of cooling of the Compton cloud on the outflow formation 
rate in an accretion disk around a black hole. We carry out a time dependent numerical simulation 
where both the hydrodynamics and the radiative transfer processes are coupled together. We consider 
a two-component accretion flow in which the Keplerian disk is immersed into an accreting low-angular momentum
flow (halo) around a black hole. The soft photons which originate from the Keplerian
disk are inverse-Comptonized by the electrons in the halo and the region
between the centrifugal pressure supported shocks and the horizon. 
We run several cases by changing the rate of the Keplerian disk and see the 
effects on the shock location and properties of the outflow and the spectrum.
We show that as a result of Comptonization of the Compton cloud, the cloud becomes cooler
with the increase in the Keplerian disk rate. As the resultant thermal pressure is reduced,
the post-shock region collapses and the outflow rate is also reduced. Since the 
hard radiation is produced from the post-shock region, and the spectral slope 
increases with the reduction of the electron temperature, the cooling 
produces softer spectrum. We thus find a direct correlation between the 
spectral states and the outflow rates of an accreting black hole.
\end{abstract}


\section{Introduction}
It is generally believed that the outflows and jets in a compact binary system containing black holes
originate from the disk itself. There are several hydrodynamical models of the formation of outflows
from the disks ranging from the twin-exhaust model of \citet{bland74}, to the self-similar
models of \citet{bland82} and \citet{bland99}. Assuming that the 
outflows are transonic in nature, \citet{fuku83} and \citet{chakra86} computed the velocity 
distribution without and with rotational motion in the flow and showed that the flow 
could become supersonic close to the black hole.
Camenzind and his group extensively worked on the magnetized jets and showed that the acceleration and collimation
of the jets could be achieved \citep[e.g.,][]{appl93}. 
In a subsequent two component transonic flow model, \citet{chakra95} pointed out
that the jets could be formed only from the inner part of the disk, which is 
also known as the Centrifugal pressure dominated boundary layer or CENBOL. The CENBOL is 
the Compton cloud in a two component flow and is the region between the accretion shock 
formed by centrifugal barrier at a few tens of Schwarzschild radii ($r_g$) and the inner sonic point
located at $\sim 2.5r_g$. If this region remains hot, it would emit hard radiation and the spectrum 
of the disk would be hard. The reverse is true if the Compton cloud (CENBOL) is cooled down
and the spectrum would be soft. 

While the general picture of the outflow formation is thus understood
and even corroborated by the radio observation of the base of the powerful jet, such as in M87
\citep{juno99} that the base of the jet is only a few tens of $r_g$,
a major question still remained: what fraction of the matter is driven out from the disk and
what are the flow parameters on which this fraction depends?  In a numerical simulation
using smoothed particle hydrodynamics (SPH),  \citet{molt94},
showed that the outflow rates from an inviscid accretion flow strongly depends on the 
outward centrifugal force and 15-20 percent matter can be driven out of the
disk. \citet{chakra98, chakra99} and \citet{chatto04} estimated the 
ratio of the outflow rate to inflow rate analytically and found that the shock 
strength determines the ratio. For very strong and very weak shocks, the outflow rates 
are very small, while for the shocks of intermediate strength, the outflow rate is significant.
These works were further refined by \citet{sing11} who 
studied the outflow rates self-consistently by modifying the Rankine-Hugoniot relations
in presence of both energy and mass dissipation. These authors showed that with the 
increase of dissipation of energy in the flow, the outflow rate is greatly reduced. This is 
in line with the recent observations \citet{fend10} that the spectrally soft states have less outflows. 

In the present paper, we concentrate on the numerical simulations of accretion flows around black holes
which are coupled to radiative transfer. While computing the time variation of the 
velocity components, density and temperature we also compute the temporal dependence of the 
spectral properties. As a result, not only we compute the outflow properties, we correlate 
them with the spectral properties -- a first in this subject. Not surprisingly,
we find that whenever the Compton cloud or the CENBOL is cooled down and the spectrum becomes
softer, the flow, originating from CENBOL, losses its drive and the outflow rate is greatly reduced.

Carrying out hydrodynamic simulations around black holes is not new. \citet{hawl84}
pioneered the study of accretion flows around black holes. Subsequent work of \citet{hawl92}
with magnetized flow indicated how the angular momentum may be transported and matter may accrete efficiently. 
These works used finite difference method.
\citet{igum96} studied two-dimensional flows but concentrated only the inner region of the
disk, namely, the region less than $20 r_g$. Their main interest was to study the transonic nature of the flow
outside the horizon. \citet{ryu95} studied unstable shocks in an inviscid flow,
while \citet{ryu97} studied the oscillations of the centrifugal pressure driven shocks 
when Rankine-Hugoniot relations are not satisfied. In a significant development which connects the time dependence
of accretion flows with quasi-periodic oscillations, \citet{molt96b} suggested that 
the shocks can oscillate whenever the cooling time scale roughly matches with the infall timescale 
in the post-shock region. Although only the bremsstrahlung cooling was used, this work provides a 
natural explanation of the quasi-periodic oscillations or QPOs which are observed in black hole candidates. 
\citet{lanz98} carried out the SPH simulation in a region with a radial extent
of $50 r_g$ in two dimensions and concentrated on the shock formation. They demonstrated 
that a high viscosity can remove the shock from an accretion flow \citep[see also,][]{giri12}.
\citet{igum98} used the finite difference
method and allowed the heat generated by viscosity to be radiated away or absorbed totally.
The computational box was up to $300r_g$, but the outer boundary condition was that of a near Keplerian
flow having no radial velocity. The inner boundary was kept at $3r_g$. Thus the possibility
of having a shock or the inner sonic point was excluded. The detailed radiative transfer, especially
Comptonization was not included. To our knowledge, ours is a first step to include Comptonization
coupled to hydrodynamic simulations where both the sources of soft photons (Keplerian disk)
and the hot electrons (low angular momentum flows) are included. Evidence of such a two-component
flow is present in many of the observations of the black hole candidates \citep{smit01, smit02, sori01, wu02}.

In the next Section, we discuss the geometry of the soft photon source 
and the Compton cloud in our Monte Carlo simulations. The variation of 
the thermodynamic quantities and other vital parameters are obtained 
inside the Keplerian disk and the Compton cloud which 
are required for the Monte Carlo simulations. In \S 3, we describe 
the simulation procedure and in \S 4, we present the
results of our simulations. Finally, in \S 5, we make concluding remarks.

\section{Geometry of the electron cloud and the soft photon source}

In Figure 1, we present the cartoon diagram of our simulation set up for 
the Compton cloud with a specific angular momentum (angular momentum per unit mass) $\lambda=1.73$.
The sub-Keplerian matter is injected from the outer boundary at $R_{in} = 100 r_g$ ($r_g=2GM_{bh}/c^2$).
The Keplerian disk resides at the equatorial plane of the cloud. The outer edge of this disk
is located at $R_{out} = 200 r_g$ and it extends up to the marginally stable orbit $R_{ms} = 3 r_g$.
At the centre, a black hole of mass $M_{bh}$ is located. The soft photons emerging out
of the Keplerian disk are intercepted and reprocessed via Compton or inverse-Compton scattering
by the sub-Keplerian matter. An injected photon may undergo no scattering at all or a single or multiple
with the hot electrons in between its emergence from the
Keplerian disk and its escape from the halo. The photons which enter the black hole are absorbed.
\subsection{Distribution of temperature and density inside the Compton cloud}

A realistic accretion disk is expected to be three-dimensional. Assuming axisymmetry, 
we have calculated the flow dynamics using a finite difference method which uses the principle of 
total variation diminishing (TVD) to carry out hydrodynamic simulations 
\citep[see,][and references therein;]{ryu97,giri10}. 
At each time step, we carry out Monte Carlo simulation to obtain the cooling/heating 
due to Comptonization. We incorporate the cooling/heating of 
each grid while executing the next time step of hydrodynamic simulation.
The numerical simulation for the two-dimensional flow has been carried out with 
$512 \times 512$ cells in a $100 r_g \times 100 r_g$ box. We choose the units in a way that
the outer boundary ($R_{in}$) is unity and the matter density at the outer boundary is
also normalized to unity. We assume the black hole to be 
non-rotating and we use the pseudo-Newtonian potential $-\frac{1}{2(r-1)}$ 
\citep{pacz80} to calculate the flow geometry around a black hole 
(Here, $r$ is in the unit of Schwarzschild radius $r_g$.). Velocities and specific angular 
momenta are measured in units of $c$, the velocity of light and $r_g c$ respectively.
In Figures (2) and (3) we show the snapshots of the density 
and temperature (in keV) profiles obtained in a steady state purely from the hydrodynamic
simulation. For both the angular momenta cases, we find the presence of two shocks in the 
accretion flow in the steady state: one paraboloidal shock and the other oblate spheroidal 
shock near the equatorial plane \citep{molt96a}. Due to the 
increased centrifugal barrier, the shock forms at a larger radius for the case $\lambda=1.76$.


\subsection{Properties of the Keplerian disk}

In our simulation, the soft photons are produced from a Keplerian disk, the inner and the outer edges of which have been kept 
fixed at the marginally stable orbit $R_{ms}$, and at $R_{out}=200 r_g$ respectively.
The source of the soft photons has a multi-color blackbody 
spectrum coming from a standard \citep[hereafter SS73]{shak73} Keplerian disk. The disk is
assumed to be optically thick and the opacity due to free-free absorption is assumed to be more 
important than the opacity due to scattering. The emission is black body type with the local surface temperature (SS73):
\begin{equation}
T(r) \approx 5 \times 10^7 (M_{bh})^{-1/2}(\dot{M_d}_{17})^{1/4} (2r)^{-3/4} \left[1- \sqrt{\frac{3}{r}}\right]^{1/4} K .
\end{equation}
The total number of photons emitted from the disk surface is obtained by integrating
over all frequencies ($\nu$) and is given by,
\begin{equation}
n_\gamma(r) = \left[16 \pi \left( \frac{k_b}{h c} \right)^3 \times 1.202057 \right]
\left(T(r)\right)^3
\end{equation}
The disk between radius $r$ to $r+\delta r$ produces $dN(r)$ number of soft photons.
\begin{equation}
dN(r) =  4 \pi r \delta r H(r) n_\gamma(r),
\end{equation}
where, $H(r)$ is the half height of the disk given by:
\begin{equation}
H(r) = 10^5 \dot{M_d}_{17} \left[1- \sqrt{\frac{3}{r}}\right] \rm{cm}.
\end{equation}
In the Monte Carlo simulation, we incorporated the directional effects of photons 
coming out of the Keplerian disk with the maximum number of photons emitted in the 
$Z$-direction and minimum number of photons are generated along the plane of the disk. 
Thus, in the absence of photon bending effects, the disk is invisible as seen 
edge on. The position of each emerging photon is   
randomized using the distribution function (Eq. 3). In the above equations, the mass 
of the black hole $M_{bh}$ is measured in units of the mass of the Sun ($M_\odot$), 
the disk accretion rate $\dot{M_d}_{17}$ is in units of $10^{17}$ gm/s. We chose 
$M_{bh} = 10$ in the rest of the paper. Generally, we follow 
\citet{ghos09, ghos10} while modeling the soft photon source. 

\section{Simulation Procedure}

For a particular simulation, we use the Keplerian disk rate ($\dot{m}_d$) and the
sub-Keplerian halo rate ($\dot{m}_h$) as parameters. The specific energy ($\epsilon$) and the specific
angular momentum ($\lambda$) determines the hydrodynamics (shock location, number 
density and velocity variations etc.) and the thermal properties of the sub-Keplerian matter.
We assume the absorbing boundary condition at $r = 1.5$ since any inward pointing photon at that radius would be
sucked into the black hole.

\subsection{Details of the hydrodynamic simulation code}

To model the initial injection of matter, we consider an axisymmetric flow of gas in the pseudo-Newtonian
gravitational field of a black hole of mass $M_{bh}$ located at the centre in the cylindrical coordinates
$[R,\theta,z]$. We assume that at infinity, the gas pressure is negligible and the
energy per unit mass vanishes. We also assume that the gravitational field
of the black hole can be described by \citet{pacz80} potential,
$$
\phi(r) = -{GM_{bh}\over(r-r_g)}, 
$$
where, $r=\sqrt{R^2+z^2}$. 
We assume a polytropic equation of state for the
accreting (or, outflowing) matter, $P=K \rho^{\gamma}$, where,
$P$ and $\rho$ are the isotropic pressure and the matter density
respectively, $\gamma$ is the adiabatic index (assumed
to be constant throughout the flow, and is related to the
polytropic index $n$ by $\gamma = 1 + 1/n$) and $K$ is related
to the specific entropy of the flow $s$. $K$ is not constant but is allowed to 
vary due to radiative processes. The details of the code is
described in \citet{molt96a} and in \citet{giri10}.

Our computational box occupies one quadrant of the R-z plane with $0 \leq R \leq 100$ and $0 \leq z \leq 100$.
The incoming gas enters the box through the outer boundary, located at $R_{in} = 100$. We have chosen the density
of the incoming gas ${\rho}_{in} = 1$ for convenience. In the absence of self-gravity
and cooling, the density is scaled out, rendering the simulation results valid for any accretion rate.
As we are considering only energy flows while keeping the boundary of the numerical grid at a finite distance, we need the sound speed $a$ (i.e., temperature) of the flow and the 
incoming velocity at the boundary points.  In order to mimic the horizon of 
the black hole at the Schwarzschild radius, we place an absorbing inner boundary
at $r = 1.5 r_g$, inside which all material is completely absorbed into the black hole. For the background matter
(required to avoid division by zero) we use a stationary gas with density ${\rho}_{bg} = 10^{-6}$ 
and sound speed (or temperature) the same as that of the incoming gas. Hence the incoming matter has a 
pressure $10^6$ times larger than that of the background matter. All
the calculations were performed with $512 \times 512$ cells, so each grid has a size of $0.19$
in units of the Schwarzschild radius.

\subsection{Details of the radiative transfer code}

To begin a Monte-Carlo simulation, we generate photons from the Keplerian disk with
randomized locations as mentioned in the earlier section. The energy of the soft photons at
radiation temperature $T(r)$ is calculated using the Planck's distribution formula, where the 
number density of photons ($n_\gamma(E)$) having an energy $E$ is expressed by \citet{pozd83},
\begin{equation}
n_\gamma(E) = \frac{1}{2 \zeta(3)} b^{3} E^{2}(e^{bE} -1 )^{-1}, 
\end{equation}
where, $b = 1/kT(r)$ and $\zeta(3) = \sum^\infty_1{l}^{-3} = 1.202$, the Riemann zeta function.
Using another set of random numbers we obtained the direction of the injected photon and with yet
another random number we obtained a target optical depth $\tau_c$ at which 
the scattering takes place. The photon was followed within the electron cloud till the optical depth ($\tau$)
reached $\tau_c$. The increase in optical depth ($d\tau$) during its traveling of
a path of length $dl$ inside the electron cloud is given by: $d\tau = \rho_n \sigma dl$, where
$\rho_n$ is the electron number density.

The total scattering cross section $\sigma$ is given by Klein-Nishina formula:
\begin{equation}
\sigma = \frac{2\pi r_{e}^{2}}{x}\left[ \left( 1 - \frac{4}{x} - \frac{8}{x^2} \right) ln\left( 1 + x \right) + \frac{1}{2} + \frac{8}{x} - \frac{1}{2\left( 1 + x \right)^2} \right],
\end{equation}
where, $x$ is given by,
\begin{equation}
x = \frac{2E}{m c^2} \gamma \left(1 - \mu \frac{v}{c} \right),
\end{equation}
$r_{e} = e^2/mc^2$ is the classical electron radius and $m$ is the mass of the electron.

We have assumed here that a photon of energy $E$ and momentum $\frac{E}{c}\bf{\widehat{\Omega}}$
is scattered by an electron of energy $\gamma mc^{2}$ and momentum 
$\overrightarrow{\bf{p}} = \gamma m \overrightarrow{\bf{v}}$, 
with $\gamma = \left( 1 - \frac{v^2}{c^2}\right)^{-1/2}$ and $\mu = \bf{\widehat{\Omega}}. \widehat{\bf{v}}$.
At this point, a scattering is allowed to take place. The photon selects an electron and the energy
exchange is computed using the Compton or inverse Compton scattering formula. The electrons
are assumed to obey relativistic Maxwell distribution inside the Compton cloud.
The number $dN(p)$ of Maxwellian electrons having momentum between
$\vec{p}$ to $\vec{p} + d\vec{p}$ is expressed by,
\begin{equation}
dN(\vec{p}) \propto exp[-(p^2c^2 + m^2c^4)^{1/2}/kT_e]d\vec{p}.
\end{equation}

In passing we wish to note that in the region of our interest (i.e. $\leq 100 r_g$) and for stellar mass black holes, 
the free-free absorption and/or emission is inefficient \citep[e.g.,][]{nara95, das08}. This can be shown as follows: 

The Compton cooling rate for a thermal distribution  of nonrelativistic electrons of
number density $n_e$ and temperature $T_e$ is \citep{rybi79, nied97}: 
$C_{c} = \frac{4kT_e}{m_e c^2}cn_e\sigma_T U_{rad}$,
($U_{rad}$ is the radiation energy density, $\sigma_T$ is the Thomson scattering
cross-section, $c$ is the speed of light, $m_e$ is the mass of the electron and $k$
is the Boltzmann constant) while the bremsstrahlung cooling rate is given by $C_{b} = 1.4 \times 10^{-27} n^2_e T_e^{1/2}$. 
Using usual formulas for $U_{rad}(r)$ and $M_{bh} = 10$, the ratio of the cooling rates turns out to be
$$
R_c=\frac{C_c}{C_b} = 1.1 \times 10^{19} \dot{m}_d r^{-3}(1-\sqrt\frac{3}{r}) \frac{T_e^{1/2}}{n_e}.
$$
To compute $R_c$, assume, $\dot{m}_h = 1.0$ and $\dot{m}_d = 1.0$. For $r \sim 100 r_g$ (pre-shock), 
$T_e \sim 10^8$ K, $n_e \sim 10^{15}$/cc, $R_c \sim 10^2$ and for $r \sim 10r_g$ (post-shock), $T_e \sim 10^9$ K, 
$n_e \sim 10^{16}$/cc $ R_c \sim 10^4$. Thus both in the pre-shock and the post-shock regions, Comptonization is much 
more effective compared to bremsstrahlung. In the simulation we ignore bremsstrahlung. Of course, 
this conclusion depends on the mass of the black hole and should be considered for super-massive black holes, for instance.

\subsubsection{Calculation of energy reduction using Monte Carlo code:}

We divide the Keplerian disk in different annuli of width $D(r)=0.5$.
Each annulus having mean radius $r$ is characterized by its average 
temperature $T(r)$. The total number of photons emitted from the disk 
surface of each annulus can be calculated using Eq. 3. 
This total number comes out to be $\sim~10^{41-42}$ per second for $\dot{m}_d = 1.0$.
In reality, one cannot inject these many number of photons in Monte Carlo simulation
because of the limitation of computation time. So we replace this large number of photons
by a lower number of bundles of photons, say, $N_{comp}(r)~\sim~10^7$ and calculate a weightage factor
$$
f_W = \frac{dN(r)}{N_{comp}(r)}.
$$
Clearly, from annulus to annulus, the number of photons in a bundle will vary. 
This is computed from the standard disk model and is used to compute the change of energy 
by Comptonization. When the injected photon is inverse-Comptonized
(or, Comptonized) by an electron in a volume element of size $dV$, 
we assume that $f_W$ number of photons has suffered similar
scattering with the electrons inside the volume element $dV$. If the energy
loss (gain) per electron in this scattering is $\Delta E$, we multiply this
amount by $f_W$ and distribute this loss (gain) among all the electrons inside
that particular volume element. This is continued for all the  $N_{comp}(r)$ bundles of photons
and the revised energy distribution is obtained.

\subsubsection{Computation of the temperature distribution after cooling} 

Since the hydrogen plasma considered here is ultra-relativistic ($\gamma=\frac{4}{3}$ throughout
the hydrodynamic simulation), thermal energy per particle is $3k_BT$ where $k_B$ is
Boltzmann constant, $T$ is the temperature of the particle.
The electrons are cooled by the inverse-Comptonization of the soft photons
emitted from the Keplerian disk. The protons are cooled because of the
Coulomb coupling with the electrons. Total number of electrons inside
any box with the centre at location $(ir,iz)$ is given by,
\begin{equation}
dN_e(ir,iz) = 4\pi rn_e(ir,iz)drdz, 
\end{equation}
where, $n_e(ir,iz)$ is the electron number density at $(ir,iz)$ location, and
$dr$ and $dz$ represent the grid size along $r$ and $z$ directions respectively. So
the total thermal energy in any box is given by $3k_BT(ir,iz)dN_e(ir,iz) = 12\pi
rk_BT(ir,iz)n_e(ir,iz)drdz,$ where $T(ir,iz)$ is the temperature at $(ir,iz)$
grid. We calculate the total energy loss (gain) $\Delta E$ of electrons inside the
box according to what is presented above and subtract that amount to get the
new temperature of the electrons inside that box as
\begin{equation}
k_BT_{new}(ir,iz) = k_BT_{old}(ir,iz)-\frac{\Delta E}{3dN_e(ir,iz)}.
\end{equation}

\subsection{Coupling procedure}

The hydrodynamic and the radiative transfer codes are coupled together following 
the same procedure as in \citet{ghos11}.
Once a quasi steady state is achieved using the non-radiative 
hydro-code, we compute the radiation spectrum using the
Monte Carlo code. This is the first approximation of the spectrum. 
To include the cooling in the coupled code, we follow these steps: 
(i) We calculate the velocity, density and temperature profiles of the electron cloud from the output of the hydro-code. 
(ii) Using the Monte Carlo code we calculate the spectrum.
(iii) Electrons are cooled (heated up) by the inverse-Compton (Compton) scattering. We calculate the
amount of heat loss (gain) by the electrons and its new temperature and energy distributions and
(iv) Taking the new temperature and energy profiles as initial condition, we run the hydro-code 
for a period of time much shorter (by a factor of ~$20$) than the cooling or infall time scale. 
Subsequently, we repeat the steps (i-iv). In this way, we see how the spectrum 
is modified and eventually converged as the iterations proceed.

Earlier we mentioned that we injected $N_{comp}(r)$ bundles. 
Before the choice of $N_{comp}(r) ~\sim~10^7$ was made, we varied $N_{comp}(r)$ from $\sim 10^3$ to $\sim 1.5\times 10^7$
and conducted a series of runs to ensure that our final result does not suffer from any statistical effects. Figure 4 shows 
the average electron temperature of the cloud when $N_{comp}(r)$ was varied. There is a clear convergence in our result
for $N_{comp}(r) > 3\times 10^6$. Thus, our choice of  $N_{comp}(r)~\sim~10^7$ is quite safe.

All the simulations are carried out assuming a stellar mass black hole $(M_{bh} = 10{M_\odot})$.
The procedures remain equally valid for massive/super-massive black holes though free-free absorption/emission may have to be included.
We carry out the simulations for more than $\sim 10$ dynamical time-scales.
In reality, this corresponds to a few seconds in physical units for the chosen gridsize.

\section{Results and Discussions}
\begin {tabular}[h]{cccc}
\multicolumn{4}{c}{Table 1: Parameters used for the simulations.}\\
\hline Case & $\epsilon, \lambda$ &  $\dot{m}_h$ & $\dot{m}_d$ \\
\hline
Ia & 0.0021, 1.76 & 1.0 & No Disk \\
Ib & 0.0021, 1.76 & 1.0 & 0.5     \\
Ic & 0.0021, 1.76 & 1.0 & 1.0     \\
Id & 0.0021, 1.76 & 1.0 & 2.0     \\
\hline
IIa & 0.0021, 1.73 & 1.0 & No Disk  \\
IIb & 0.0021, 1.73 & 1.0 & 0.5      \\
IIc & 0.0021, 1.73 & 1.0 & 1.0      \\
IId & 0.0021, 1.73 & 1.0 & 2.0      \\
\hline
\end{tabular} \\

In Table 1, we list various Cases with all the simulation parameters used 
in the present paper. The specific energy ($\epsilon$) and specific angular momentum ($\lambda$) 
of the sub-Keplerian halo are given in Column 2. Columns 3 and 4 give the halo ($\dot{m}_h$) and 
the disk ($\dot{m}_d$) accretion rates. The corresponding cases are marked in Column 1. 
In Cases Ia and IIa, no Keplerian disk was placed in the 
equatorial plane of the halo. These are non-radiative hydrodynamical 
simulations and no Compton cooing is included. To show the 
effects of Compton cooling on the hydrodynamics of the flow, the Cases I(b-d) and II(b-d) are 
run for the same time as the Cases Ia and IIa.

\subsection{Properties of the shocks in presence of cooling}

In Figures 5(a) and 5(b), we present the time variation of the shock location (in units of $r_g$)
for various Cases (marked on each curve) given in Table 1. All the solutions 
exhibit oscillatory shocks.  For no cooling, the higher angular momentum produces shocks at a higher
radius, which is understandable, since the shock is primarily centrifugal force
supported. However, as the cooling is increased the average shock location decreases 
since the cooling reduces the post-shock thermal pressure and the shock could not be 
sustained till higher thermal pressure is achieved at a smaller radius. 
The corresponding oscillations are also suppressed. The average shock location 
is found to be almost independent of the specific angular momentum at this stage. This is because 
the post-shock region is hotter in Cases I(a-d), and thus it cools at a very smaller time scale.
In Figure 6, we show the colour map of the temperature distribution at the end of our
simulation. We zoomed the region $50 r_g \times 50 r_g$. The specific angular momentum is
$1.76$ in the left panel and $1.73$ in the right panel. Cases are marked. We note the 
collapse of the post-shock region as $\dot{m}_d$ is increased gradually. We take the 
post-shock region in each of these cases, and plot in Figures 7(a) and 7(b)
the average temperatures of the post-shock region only for those cases where the 
cooling due to Comptonization is included. The average temperature was obtained by the 
optical depth weighted averaging procedure prescribed in \citet{chakra95}.
The average temperature in the post-shock region is reduced rapidly as the supply 
of the soft photons is increased. 

\subsection{Effects of Comptonization on the outflow rates}

We now concentrate on how the outflow rate is affected by the Comptonization. Outflows
move to very large distances and thus must not be bound to the system, i.e., the 
specific energy should be positive. Matter should also be of higher entropy 
as it is likely to be relativistic. Because of this we wish to concentrate on the 
behaviour of matter which have highest energy and entropy. Though we injected matter
at the outer edge with a constant specific energy, the energy of matter in the post-shock region
is redistributed due to turbulence, Compton cooling and shock heating. Some entropy is generated as well.
The high energy and high entropy matter escape in the form of a hollow cone around the axis. 
It is thus expected that if the post-shock region itself is collapsed due to 
Comptonization, the outflows will also be quenched. We show this effect in our result. In Figure 8,
we present the specific energy distribution for both the specific angular momenta (left panel 
for $\lambda=1.76$ and right panel for $\lambda=1.73$) for all the cooling Cases (marked in each box)
at the end of our simulation. The velocity vectors are also plotted. 
The scale on the right gives the specific energy. First we note that the jets are stronger
for higher angular momentum. This is because the post-shock region (between the shock and the 
inner sonic point close to the horizon) is hotter. Second, lesser and lesser amount of 
matter has higher energy as the cooling is increased. A similar observation could be made from 
Figure 9, where the entropy distribution is plotted. The jet matter having upward pointing vectors
have higher entropy. However, this region shrinks with the increase in Keplerian rate, as the cooling 
becomes significant the outward thermal drive is lost. Here, the velocity vectors are of length $0.05$ at the
outer boundary on the right  and others are scaled accordingly. 

In order to quantify the decrease in the outflow rates with cooling, we define two types of outflow rates.
One is $\dot{M}_{out}$ which is defined to be the rate at which outward pointing flow 
leaves the computational grid. This will include both the high and low energy components of the 
flow. In Figures 10(a) and (b), we show the results of time variation of the ratio $R_{\dot{m}}$ 
(${\dot{M}}_{out}$ / ${\dot{M}}_{in}$) for the four cases (marked in each box), ${\dot{M}}_{in}$ 
being the constant injection rate on the right boundary. While the ratio is  
clearly a time varying quantity, we observe that with the increase in cooling the 
ratio is dramatically reduced and indeed become almost saturated as soon as some 
cooling is introduced. Our rigorous finding once again verified what was
long claimed to be the case, namely, the spectrally soft states
(those having a relatively high Keplerian rate) have weaker jets because of the
presence of weaker shocks \citep{chakra99, das01}. 

Another measure of the outflow rate would be to concentrate only those matter which have 
high positive energy and high entropy. For concreteness, we concentrate only on matter
outflowing within $x=20r_g$ at the upper boundary of our computational grid. We define this to be
$J_{\dot{m}}$ ($= \frac{\dot{M}_{jet}}{\dot{M}_{in}}$). Figures 11(a) and 11(b) show time variation of 
$J_{\dot{m}}$. The different cases are marked on the curves. 
This outflow rate fluctuates with time. We easily find that the cooling process reduces 
this high energy component of matter drastically. Thus both the slow moving outflows and fast 
moving jets are affected by the Comptonization process at the base. 

\subsection{Spectral properties of the Disk-Jet system}

In each simulation we also store the photons emerging out of the Computational grid
after exchanging energy and momentum with the free electrons in the disk matter. 
When the Keplerian disk rate is increased, the injected soft photons go up, cooling every electron
in the sub-Keplerian halo component. Thus, the relative availability of the soft photons and the 
hot electrons in the disk and the jet dictates whether the emergent photons would be 
spectrally soft or hard. In Figures 12(a) and 12(b), we show three spectra for each of the specific
angular momenta: (a) $\lambda=1.76$ and (b) $\lambda=1.73$. The Cases are marked. 
We see that a spectrum is essentially made up of the soft bump (injected multi-colour black body spectrum
from the Keplerian disk), and a Comptonized spectrum with an exponential cutoff -- the cutoff energy being
dictated by the electron cloud temperature. If we define the energy spectral index  $\alpha$ to be
$I(E) \propto E^{-\alpha}$ in the region $1-10$ keV, we note that $\alpha$ increases, i.e., the spectrum
softens with the increase in $\dot{m}_d$. This is consistent with the static model of \citet{chakra95}.

In reality, since the disk is not stationary, the spectrum also varies with time, and so is $\alpha$. In Figure 13,
we present the time variation of the spectral index for the different Cases. We draw the running mean through these
variations.  We clearly see that the spectral index goes up with the increase in the disk accretion rate.
Thus, on an average, the spectrum softens. We also find that the spectrum oscillates
quasi-periodically and the frequency is higher for higher cooling rate. This agrees with the general observations
that the QPO frequency rises with luminosity \citep{cui99, sobc00, macl09}.

It is interesting to understand the physical reason behind the shock oscillations. 
It has been argued by numerical simulation \citep{molt96b, chakra04} this oscillation is due to
a resonance effect. When the cooling time scale {\it roughly} agrees with the infall time scale, the
post-shock region, i.e., the CENBOL may oscillate. As the CENBOL oscillates, the high energy
part of the spectrum also oscillates since the CENBOL is mainly responsible to energize the soft photons 
coming out from the Keplerian disk to high energy. Indeed, when we compute the cooling time scale and 
the infall time scale, we find that in all the cases when oscillations occur, the ratio of cooling time 
scale and infall time scale
on an average remains within forty percent of each other. Of course, in the absence of cooling effects,
some times the shock oscillations is possible when the Rankine-Hugoniot condition is not fulfilled
\citep{ryu97}. The amplitude of oscillation in this case is generally quite large and the result strongly 
depends on the specific angular momentum and specific energy of the flow. Observationally, different types
of QPOs have been observed. This aspect requires a further study which is beyond the scope of this paper.

\section{Concluding remarks}

In this paper, we compute the effects of Compton cooling on the formation and rates
of the outflows and jets. Most importantly, we study the variation of the spectral slope. We employ 
a hydrodynamic code based on the total variation diminishing method coupled to a Monte Carlo
simulation to take care of the time dependent radiative transfer problem. We found many important
results which can be briefly summarized as follows: (a) The increase in the Keplerian 
disk rate cools down the post-shock region of a sub-Keplerian flow and as a result the shock 
advances closer to the black hole. (b) The cooler and smaller sized post-shock region produced
a weaker outflow/jet. (c) The spectrum becomes softer when the cooling is increased. (d) Quasi-periodic 
oscillations are produced as a result of near resonance between the cooing time scale as the infall time scale.

We coupled the hydrodynamics and Comptonization using a two step process. This is natural:
Comptonization is strongly  non-local process and the results are obtained by Monte Carlo method. However,
the hydrodynamics code is a finite difference code. So at every instant both effects cannot be taken into account. 
The effects of radiative cooling or heating due to Comptonization is included in the hydrodynamics after each 
Monte Carlo step. After that the hydrodynamic code is advanced for a time period 
which is a factor of $\sim  20$ lower than the cooling or hydrodynamic time scale, whichever is lesser 
before running the Monte-Carlo step once more. Thus our dynamics is affected by the radiative cooling or heating effects. 

In this context, it is important to note that the effects of radiation is not the same as in a single 
component flow.  In a single component disk, such as the  
standard accretion disk, the radiation field is strong (especially when the accretion rate
crosses the Eddington rate) as has been shown for example, by \citet{ohsu11}.
However, in our case, the radiation field is important only if it interacted with electrons.
In a two component flow, there are two effects which affect the dynamics of the electrons
(a) the thermal acceleration $F_{th}$ depends on the gradient of thermal pressure
and (b) the radiative acceleration $F_{rad}$ depends on the number of Compton scattering
and the amount of energy and momentum exchanged in the process (i.e., depends on local temperature).
Both $F_{th}$ and $F_{rad}$ have strong dependence on both the accretion rates.
For a fixed halo rate, $F_{rad}$ monotonically rises with disk rate but not exactly as in
\citet{ohsu11}, since the results depend on the optical depth of the
halo. $F_{th}$ need not rise monotonically, however. We find that with the rise of the disk rate,
initial rising trend of $F_{th}$ is reversed because the electrons are cooled down 
at a higher Keplerian disk rate, so the local temperature becomes low. We find that the 
ratio $F_{rad}/F_{th}$ remains much less than unity even for ${\dot m}_{d} \sim 10$.

Some of these results, as such, have been anticipated both theoretically and observationally. From the
theoretical point of view \citep{chakra95}, in the two component model showed 
that the spectrum becomes softer. Similarly, \citet{chakra99}
found that the outflow rate becomes smaller when shock becomes weaker. The shock oscillation model
observations reported by \citet{gall04, fend04} point to the
same correlation between the spectral states and the outflow rates. However, it is the first time
that we demonstrate the results with a complete coupled numerical simulations. The results we 
found are quite general and should be valid even for massive black holes, provided the 
cooling plays a role in the dynamics of the flow.

\acknowledgments
The work of HG was supported by a RESPOND grant from ISRO. 

{}

\clearpage

\vfil\eject
\begin{figure}
\centering{
\includegraphics[height=9truecm,width=9truecm,angle=0]{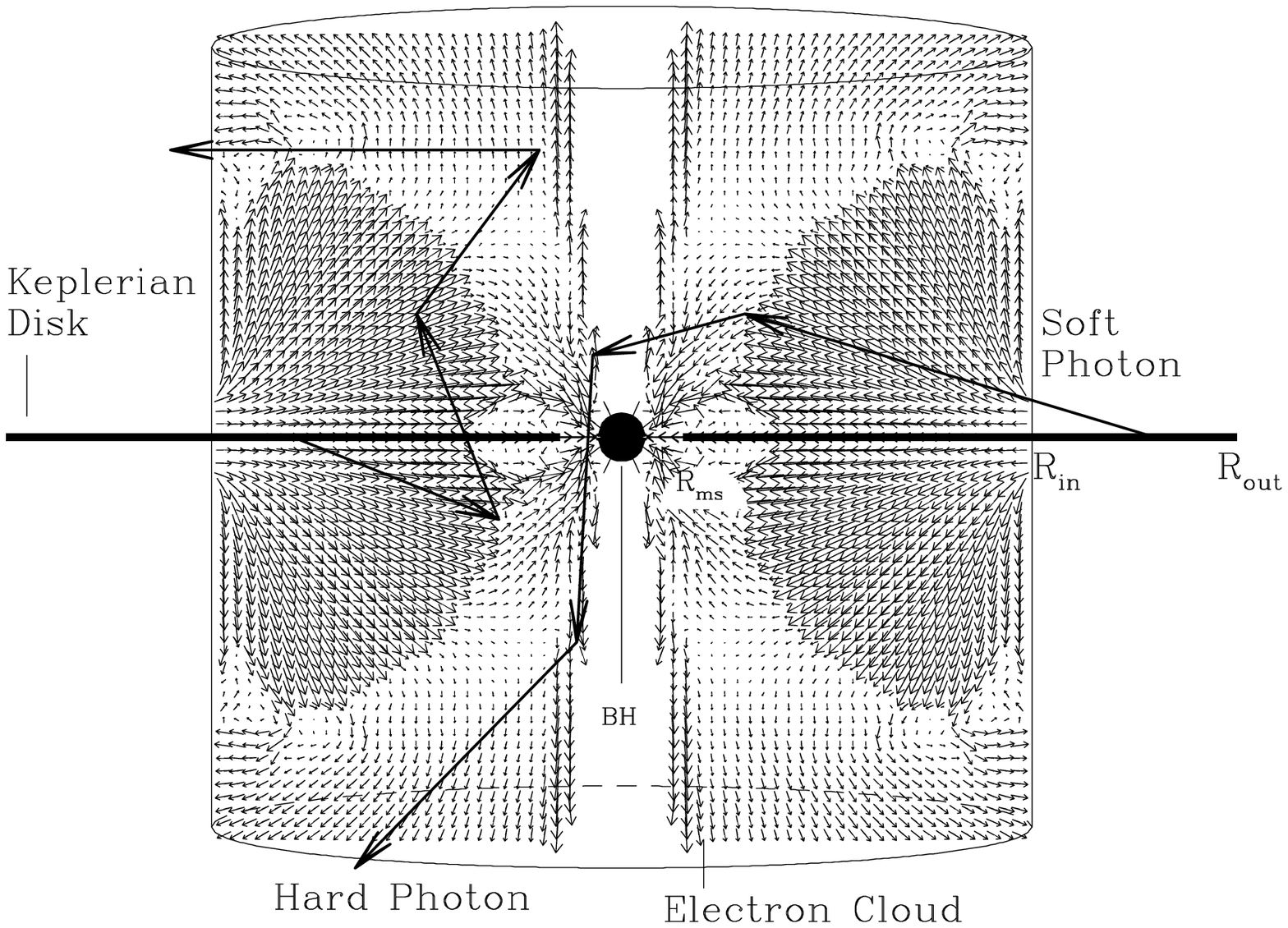}}
\caption{Schematic diagram of the geometry of our Monte Carlo simulations.  
Zigzag trajectories are the typical paths followed by the photons. The velocity
vectors of the infalling matter inside the cloud are shown. The velocity vectors 
are plotted for $\lambda=1.73$.}
\end{figure}

\vfil\eject
\begin{figure}
\centering{
\includegraphics[height=4.5truecm,width=9truecm,angle=0]{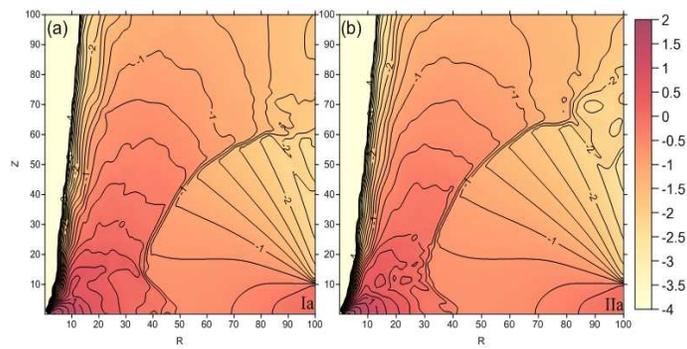}}
\caption{Logarithmic density (normalized unit) contours inside the halo for two different angular 
momenta (a) 1.76 and (b) 1.73. Logarithmic density contour levels are -4(0.25)2.}
\end{figure}

\vfil\eject
\begin{figure}
\centering{
\includegraphics[height=4.5truecm,width=9truecm,angle=0]{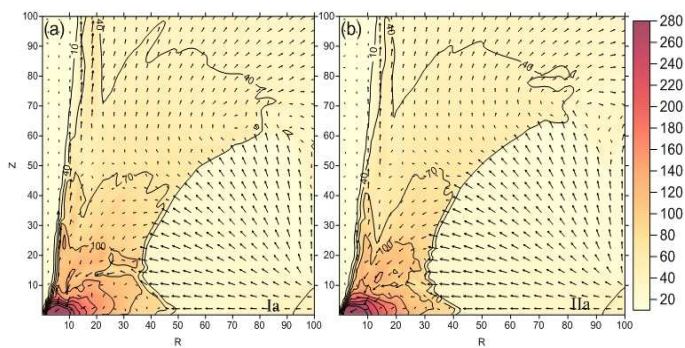}}
\caption{Temperature contours inside the halo. Velocity vectors are also provided.
Here, temperatures are in keV. Temperature contours are 10(20)280.}
\end{figure}

\vfil\eject
\begin{figure}
\centering{
\includegraphics[height=6cm,width=6cm]{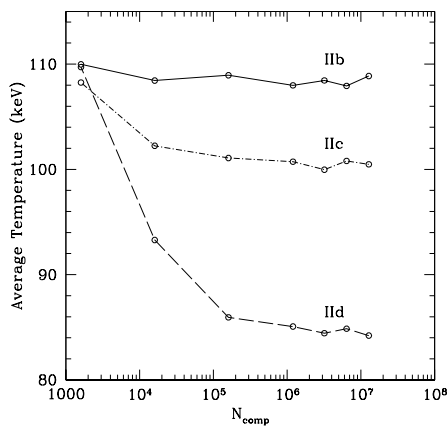}}
\caption{Variation of the average temperature (keV) of the post shock region with 
bundle of photons $N_{comp}$ for different Keplerian disk rates $\dot{m}_d$, keeping 
the halo rate fixed at $\dot{m}_h = 1.0$. 
Simulation cases (Table 1) are marked on each curve.
Clearly, the temperature converges for $N_{comp}(r) > 3\times 10^6$.
}
\end{figure}

\vfil\eject
\begin{figure}
\begin{center}
\includegraphics[width=8cm,height=8cm]{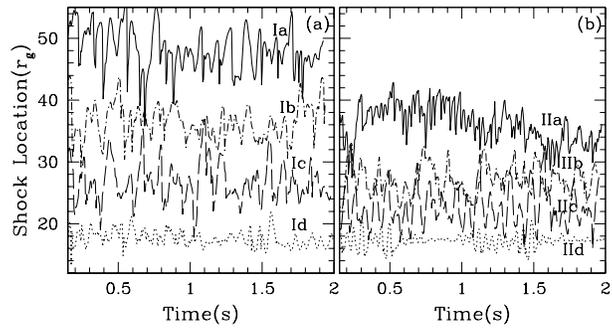}
\caption{The variation of shock location (in $r_g$) at the equatorial plane with time (in sec)
for different Keplerian disk rates $\dot{m}_d$, keeping the halo rate fixed
at $\dot{m}_h = 1.0$. Simulation Cases are marked on each curve.
(a) $\lambda = 1.76$ and (b) $\lambda = 1.73$. Cooling decreases the average shock location.}
\end{center}
\end{figure}

\vfil\eject
\begin{figure}
\begin{center}
\includegraphics[width=8cm,height=8cm]{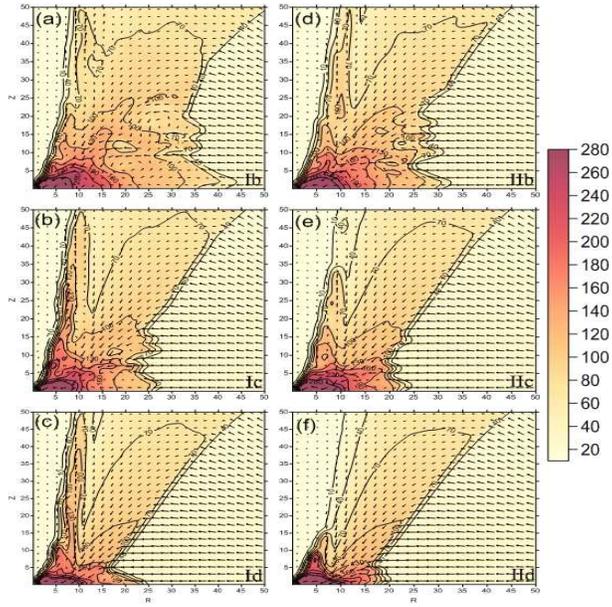}
\caption{Colour map of final temperature distributions in the region ($50 r_g \times 50 r_g$) of the 
accretion disk for different disk rates are shown. The left panel is for $\lambda = 1.76$
and the right is for $\lambda = 1.73$. As $\dot{m}_d$ is increased we find that, the high temperature 
region (dark zone, dark red online) shrinks.}
\end{center}
\end{figure}

\vfil\eject
\begin{figure}
\begin{center}
\includegraphics[width=8cm,height=8cm]{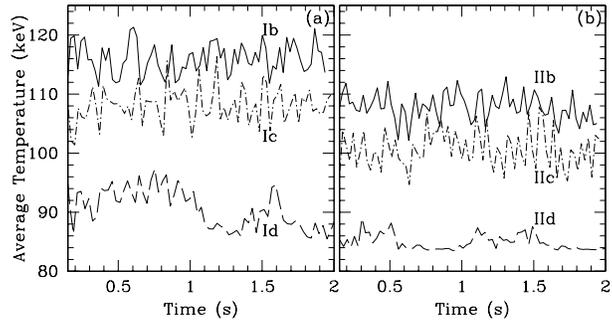}
\caption{Variation of the average temperature (keV) of the post shock region with time (sec)
for different Keplerian disk rates $\dot{m}_d$, keeping the halo rate fixed
at $\dot{m}_h = 1.0$. Parameters are the same as in Figures 5a and 5b.}
\end{center}
\end{figure}

\vfil\eject
\begin{figure}
\begin{center}
\includegraphics[width=8cm,height=8cm]{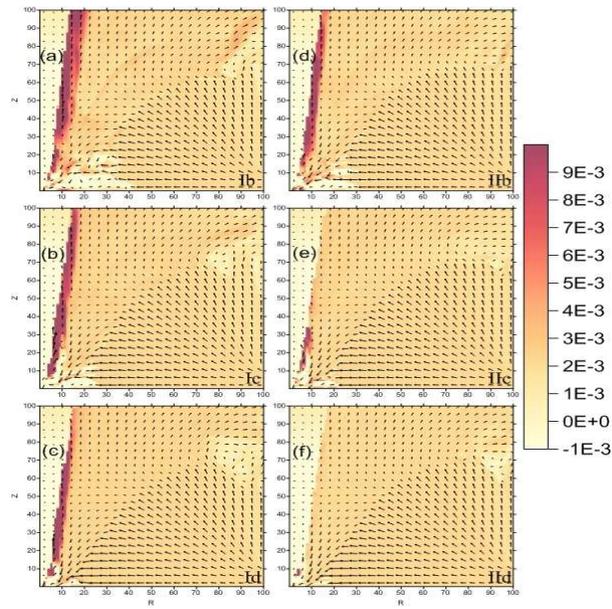}
\caption{Colour map of final specific energy distribution inside the 
accretion disk for different disk rates. The high energy matter (dark zone, dark red online) 
are ejected outward as a hollow jet. The matter with a high energy flow decreases with the
increase in disk rate. Velocity vectors at the injection boundary on the  right is of length $0.05$.}
\end{center}
\end{figure}

\vfil\eject
\begin{figure}
\begin{center}
\includegraphics[width=8cm,height=8cm]{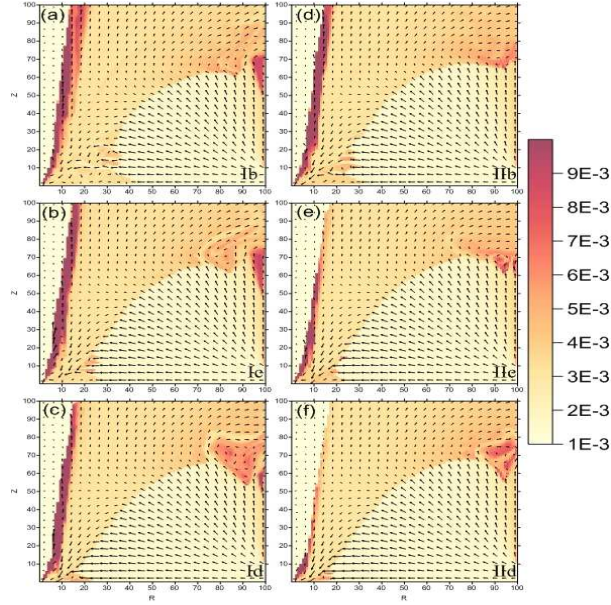}
\caption{Color map of the final entropy ($K = \frac{P}{\rho^{\gamma}}$) distribution. 
Other parameters are as in Figure 8. The high entropy flow decreases as the disk rate increases.}
\end{center}
\end{figure}

\vfil\eject
\begin{figure}
\begin{center}
\includegraphics[width=4cm,height=4cm]{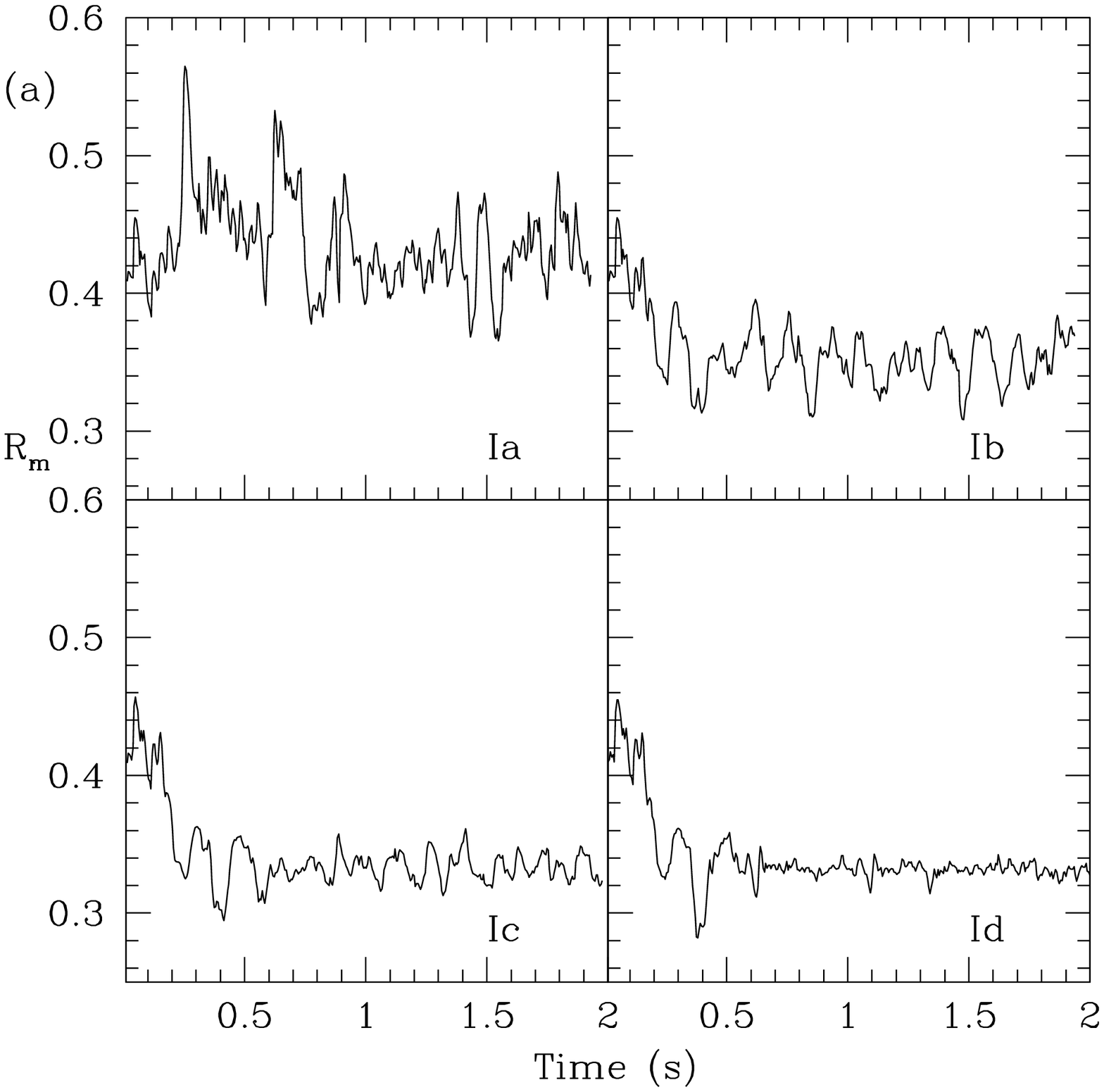}
\includegraphics[width=4cm,height=4cm]{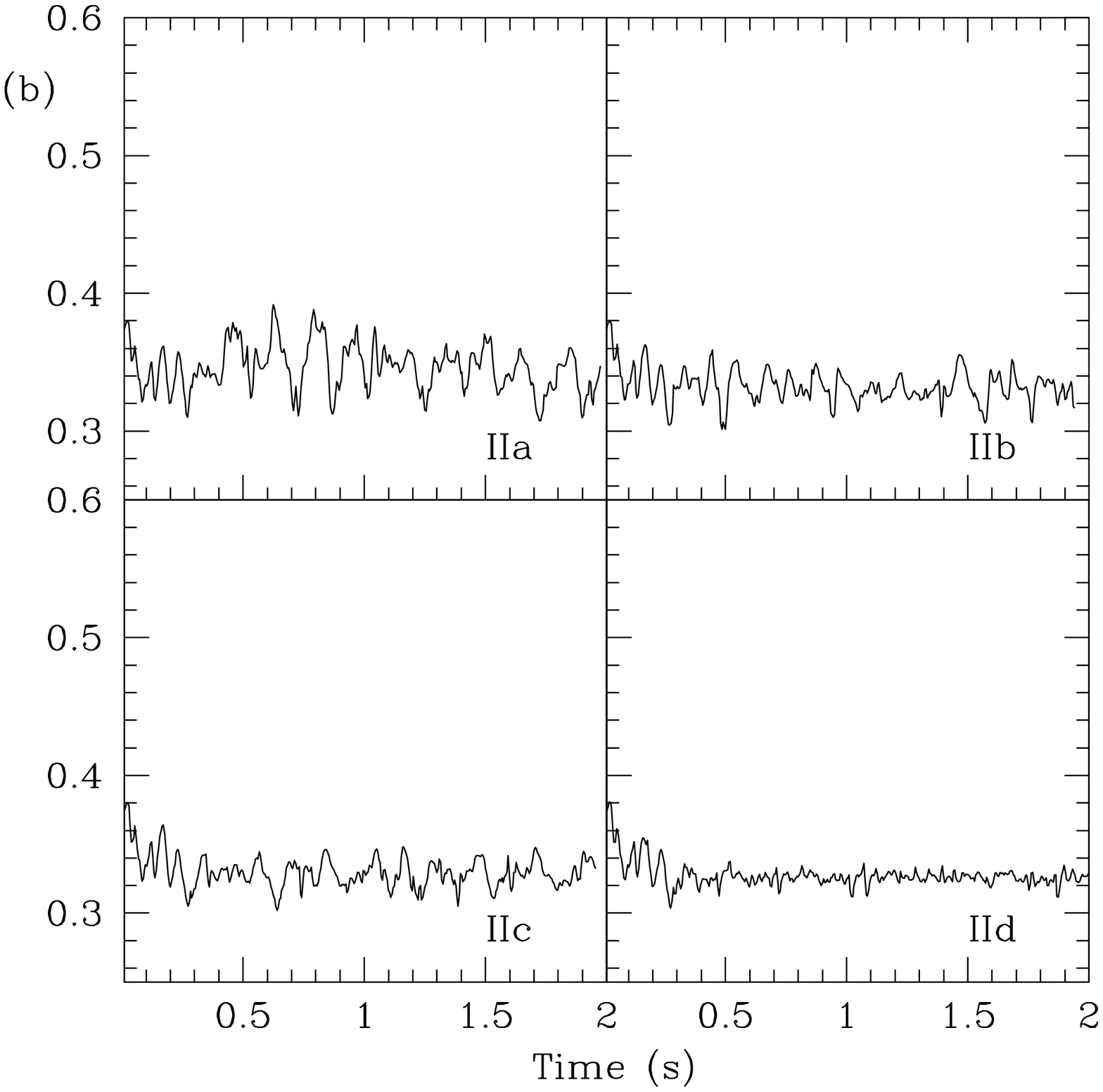}
\caption{Variations of $R_{\dot{m}} (=\frac{\dot{M}_{out}}{\dot{M}_{in}})$ with time for
different $\dot{m_d}$ is shown here. (a) $\lambda = 1.76$ and (b) $\lambda = 1.73$.
The Cases are marked in each panel. The outflow rate is the lowest for the 
highest Keplerian disk accretion rate (Cases are Id and IId).} 
\end{center}
\end{figure}

\vfil\eject
\begin{figure}
\begin{center}
\includegraphics[width=8cm,height=8cm]{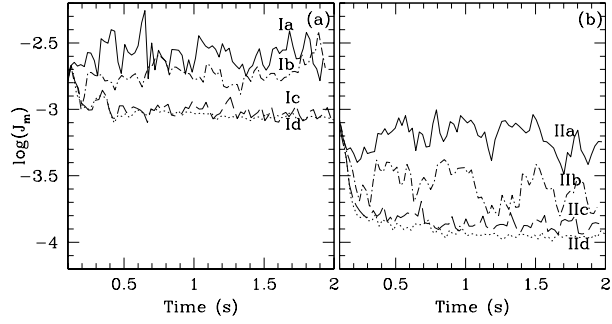}
\caption{Variations of $J_{\dot{m}}$ ($= \frac{\dot{M}_{jet}}{\dot{M}_{in}}$) with time for
different $\dot{m_d}$ is shown here. Here, $\dot{M}_{jet}$ and $\dot{M}_{in}$ are the
high entropy (also high energy) outflow and inflow rates, respectively. The left panel 
is for $\lambda = 1.76$ and the right panel is for $\lambda = 1.73$.  The Cases are marked in each curve.}
\end{center}
\end{figure}

\vfil\eject
\begin{figure} 
\begin{center}
\includegraphics[width=8cm,height=4cm]{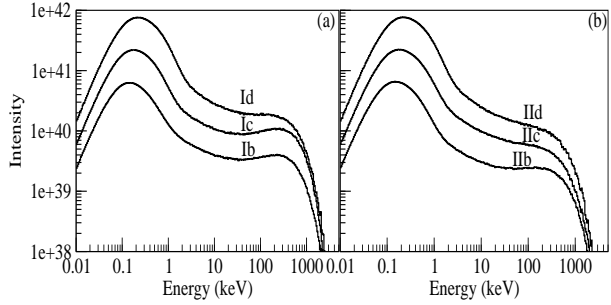}
\caption{The final emitted spectra for different disk rates are shown for (a) $\lambda = 1.76$ and (b) $1.73$. 
Corresponding Cases are marked on each curve. The spectrum appears to become
softer with the increase in $\dot{m_d}$.}
\end{center}
\end{figure}

\vfil\eject
\begin{figure}
\begin{center}
\includegraphics[width=8cm,height=4cm]{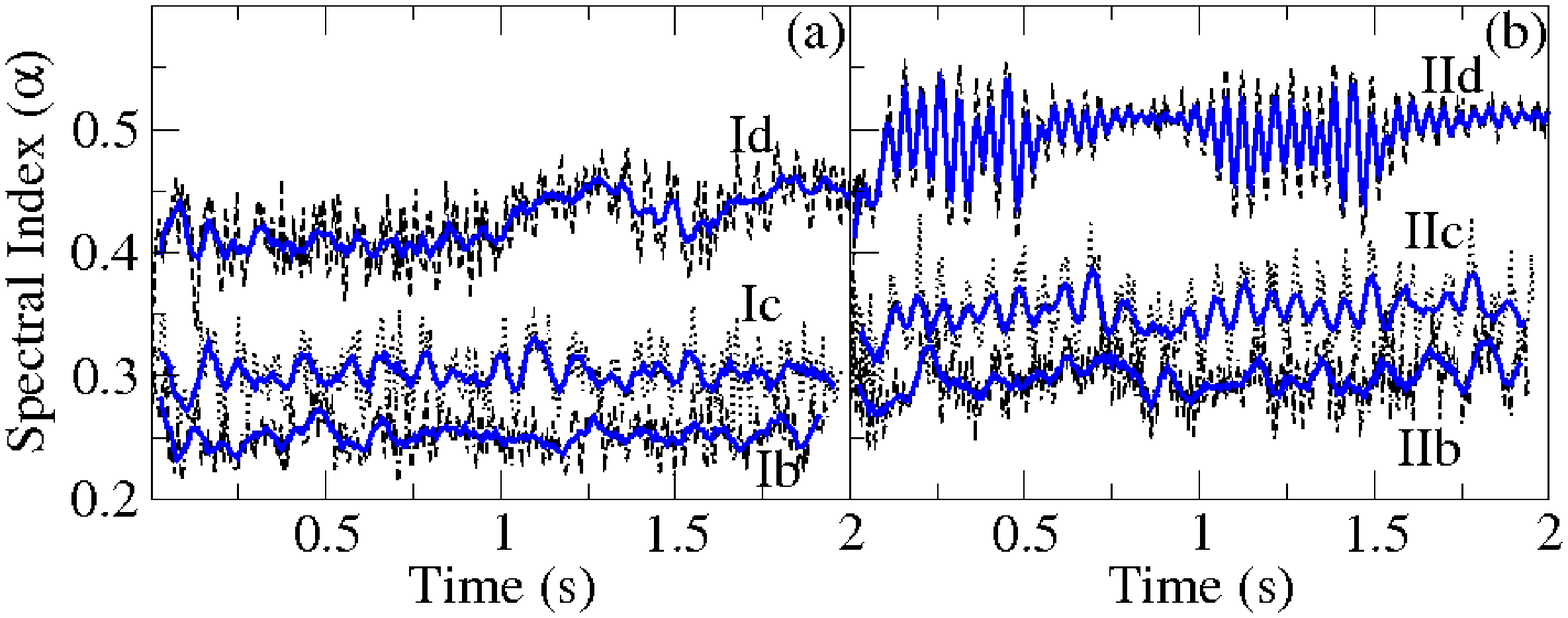}
\caption{Time variation of the spectral slope ($\alpha$, $I(E) \propto E^{-\alpha}$) for different 
disk rates and their running averages (solid line, blue online) are shown. Different Cases are
marked. We note that as the accretion rate goes up, the average $\alpha$ increases, i.e., the 
spectrum softens. The running mean gives an idea that there is a quasi-periodic variation of the 
spectral index -- the frequency is higher for higher cooling rate.}
\end{center}
\end{figure}


\begin{thebibliography}{}
\def\ref#1\par{\parshape=2 0in 14.5cm 1cm 13.5cm {#1} \par}
\parskip=0pt
\parindent=0pt

\bibitem[Appl \& Camenzind (1993)]{appl93} Appl, S., \& Camenzind, M. 1993, A\&A, 270, 71
\bibitem[Blandford \& Begelman (1999)]{bland99} Blandford, R. D., \& Begelman, M. C. 1999, MNRAS, 303, L1
\bibitem[Blandford \& Payne (1982)]{bland82} Blandford, R. D., \& Payne, D. G. 1982, MNRAS, 199, 883
\bibitem[Blandford \& Rees (1974)]{bland74} Blandford, R. D., \& Rees, M. J. 1974, MNRAS, 169, 395
\bibitem[Chakrabarti (1986)]{chakra86} Chakrabarti, S. K. 1986, ApJ, 303, 582
\bibitem[Chakrabarti (1998)]{chakra98} Chakrabarti, S. K. 1998, arXiv:astro-ph/9801079
\bibitem[Chakrabarti (1999)]{chakra99} Chakrabarti, S. K. 1999, A\&A, 351, 185
\bibitem[Chakrabarti et al. (2004)]{chakra04} Chakrabarti, S. K., Acharyya, K., \& Molteni, D. 2004, A\&A, 421, 1
\bibitem[Chakrabarti \& Titarchuk (1995)]{chakra95} Chakrabarti, S. K., \& Titarchuk, L.G. 1995, ApJ, 455, 623
\bibitem[Chattopadhyay et al. (2004)]{chatto04} Chattopadhyay, I., Das, S., \& Chakrabarti, S. K. 2004, MNRAS, 348, 846
\bibitem[Cui et al. (1999)]{cui99} Cui, W., Zhang, S. N., Chen, W., et al. 1999, ApJ, 512, L43
\bibitem[Das \& Chattopadhyay (2008)]{das08} Das, S. \& Chattopadhyay, I. 2008, New Astron., 13, 549
\bibitem[Das et al. (2001)]{das01} Das, S., Chattopadhyay, I., Nandi, A., \& Chakrabarti, S. K. 2001, A\&A, 379, 683
\bibitem[Fender et al. (2004)]{fend04} Fender, R. P., Belloni, T. M., \& Gallo, E. 2004, MNRAS, 355, 1105 
\bibitem[Fender et al. (2010)]{fend10} Fender, R. P., Gallo, E., \& Russell, D. 2010, MNRAS, 406, 1425 
\bibitem[Fukue (1983)]{fuku83} Fukue, J. 1983, PASJ, 35, 539
\bibitem[Gallo et al. (2004)]{gall04} Gallo, E., Fender, R. P., \& Pooley, G. 2004, NuPhS, 132, 363
\bibitem[Ghosh et al. (2009)]{ghos09} Ghosh, H., Chakrabarti, S. K., \& Laurent, P. 2009, IJMPD, 18, 1693
\bibitem[Ghosh et al. (2010)]{ghos10} Ghosh, H., Garain, S. K., Chakrabarti, S. K., \& Laurent, P. 2010, IJMPD, 19, 607
\bibitem[Ghosh et al. (2011)]{ghos11} Ghosh, H., Garain, S. K., Giri, K., \& Chakrabarti, S. K. 2011, MNRAS, 416, 959
\bibitem[Giri et al. (2010)]{giri10} Giri, K., Chakrabarti, S. K., Samanta, M. M., \& Ryu, D. 2010, MNRAS, 403, 516
\bibitem[Giri \& Chakrabarti (2012)]{giri12} Giri, K., \& Chakrabarti, S. K. 2012, MNRAS, 421, 666 
\bibitem[Hawley \& Balbus (1992)]{hawl92} Hawley, J. F., \& Balbus, S. A. 1992, ApJ, 400, 595 
\bibitem[Hawley et al. (1984)]{hawl84} Hawley, J. F., Smarr, L. L., \& Wilson, J. R. 1984, ApJS, 55, 211 
\bibitem[Igumenshchev et al. (1998)]{igum98} Igumenshchev, I. V., Abramowicz, M. A., \& Novikov, I. D. 1998, MNRAS, 298, 1069
\bibitem[Igumenshchev et al. (1996)]{igum96} Igumenshchev, I. V., Chen, X., \& Abramowicz, M. A. 1996, MNRAS, 278, 236
\bibitem[Junor et al. (1999)]{juno99} Junor, W., Biretta, J. A. \& Livio, M. 1999, Nature, 401, 891
\bibitem[Lanzafame et al. (1998)]{lanz98} Lanzafame, G., Molteni, D., \& Chakrabarti, S. K. 1998, MNRAS 299 799
\bibitem[McClintock et al. (2009)]{macl09} McClintock, J. E., Remillard, R. A., Rupen, M. P., et al. 2009, ApJ, 698, 1398
\bibitem[Molteni et al. (1994)]{molt94} Molteni, D., Lanzafame, G., \& Chakrabarti, S. K. 1994, ApJ, 425, 161
\bibitem[Molteni et al. (1996)]{molt96a} Molteni, D., Ryu, D., \& Chakrabarti, S. K. 1996, ApJ, 470, 460
\bibitem[Molteni et al. (1996)]{molt96b} Molteni, D., Sponholz, H., \& Chakrabarti, S. K. 1996, ApJ, 457, 805
\bibitem[Narayan \& Yi (1995)]{nara95} Narayan, R. \& Yi, I. 1995, ApJ, 452, 710
\bibitem[Niedzwiecki et al. (1997)]{nied97} Niedzwiecki, A. M., Krolik, J. H., \& Zdziarski, A. 1997, ApJ, 483, 111
\bibitem[Ohsuga \& Mineshige (2011)]{ohsu11} Ohsuga, K., \& Mineshige, S. 2011, ApJ, 736, 2
\bibitem[Paczy\'nski \& Wiita (1980)]{pacz80} Paczy\'nski, B., \& Wiita, P. J. 1980, A\&A, 88, 23
\bibitem[Pozdnyakov et al. (1983)]{pozd83} Pozdnyakov, A., Sobol, I. M., \& Sunyaev, R. A. 1983, Astrophys. Space Sci. Rev., 2, 189
\bibitem[Rybicki \& Lightman (1979)]{rybi79} Rybicki, G. B., \& Lightman, A. P. 1979, Radiative Processes in Astrophysics (John Wiley \& Sons, New York)
\bibitem[Ryu et al. (1995)]{ryu95} Ryu, D., Brown, G. L., Ostriker, J. P., \& Loeb, A. 1995, ApJ, 452, 364
\bibitem[Ryu et al. (1997)]{ryu97} Ryu, D., Chakrabarti, S. K., \& Molteni, D. 1997, ApJ, 474, 378
\bibitem[Shakura \& Sunyaev (1973)]{shak73} Shakura, N. I., \& Sunyaev, R. A. 1973, A\&A, 24, 337
\bibitem[Singh \& Chakrabarti (2011)]{sing11} Singh, C. B., \& Chakrabarti, S. K. 2011, MNRAS, 410, 2414
\bibitem[Smith et al. (2001)]{smit01} Smith, D. M., Heindl, W. A., Markwardt, C. B., \& Swank, J. H. 2001, ApJ, 554, L41
\bibitem[Smith et al. (2002)]{smit02} Smith, D. M., Heindl, W. A., \& Swank, J. H. 2002, ApJ, 569, 362
\bibitem[Sobczak et al. (2000)]{sobc00} Sobczak, G. J., Mcclintock, J. E., Remillard, R. A., et al. 2000, ApJ, 531, 537
\bibitem[Soria et al. (2001)]{sori01} Soria, R., Wu, K., Hannikainen, D., McMollough, M. \& Hunstead, R. 2001, Proceedings of a joint workshop held by the Center for Astrophysics (Johns Hopkins University) and the Laboratory for High Energy Astrophysics (NASA/ Goddard Space Flight Center) in Baltimore, MD; Eds.: T. Yaqoob and J. H. Krolik
\bibitem[Wu et al. (2002)]{wu02} Wu, K., Soria, R., \& Campbell-Wilson, D., et al. 2002, ApJ, 565, 1161
	
 

\end{thebibliography}
\end{document}